\documentclass[twoside]{book}% Specify document type
 \makeatletter
\ifx\UNDEF\mail\def\mail{ }\else\fi
\ifx\UNDEF\prange\def\prange{0 0}\else\fi

\gdef\@empty{}
\def\Mail#1 #2 {\gdef\thecontact{#1}\gdef\theaddr{#2}}
\def\Range#1 #2 {\gdef\thefirstpage{#1}\gdef\thelastpage{#2}}
{\let\'\mail \expandafter\Mail\' }  % do not remove space between ' and }
{\let\'\prange \expandafter\Range\' }   % do not remove space between ' and }
 \gdef\@shtitle{\relax}
 \long\def\shtitle#1{\gdef\@shtitle{#1}}
 \long\def\author#1{\gdef\@author{#1}}
 \def\affil#1{\par\noindent{\rm#1\par}}
 \gdef\@abstract{}
 \long\def\abstract#1{\gdef\@abstract{#1}}

 \def\maketitle{\thispagestyle{empty}\chapter{\@title}}
 \renewcommand\chapter{\if@openright\cleardoublepage\else\clearpage\fi
                    \thispagestyle{empty}%
                    \global\@topnum\z@
                    \@afterindentfalse
                    \secdef\@chapter\@schapter}
 \def\@makechapterhead#1{%
  \vspace*{50\p@}%
  {\parindent \z@ \raggedleft \normalfont
    \ifnum \c@secnumdepth >\m@ne
      \if@mainmatter
        \par\nobreak
        \vskip 20\p@
      \fi
    \fi
    \interlinepenalty\@M
    \Huge \bfseries #1\par\nobreak
    \vskip.25in
    \large\bfseries\@author\par\nobreak
    \vskip 40\p@}
    \ifx\@abstract\@empty\else{\small\@abstract\par\vskip20\p@}\fi
  }

\DeclareRobustCommand\em
        {\@nomath\em \ifdim \fontdimen\@ne\font >\z@
                       \upshape \else \slshape \fi}

\def\@begintheorem#1#2{\sl \trivlist \item[\hskip \labelsep{\bf #1\ #2}]}
\def\@opargbegintheorem#1#2#3{\sl \trivlist
     \item[\hskip \labelsep{\bf #1\ #2\ (#3)}]}

  \def\@arabic#1{\number #1} % my redefinition

\long\def\@makecaption#1#2{
    \vskip\abovecaptionskip
    \sbox\@tempboxa{{\small {\bf #1}: #2}}%
    \ifdim\wd\@tempboxa>\hsize
        {\small {\bf #1}: #2\par}
    \else
       \global\@minipagefalse
       \hbox to\hsize{\hfil\box\@tempboxa\hfil}
    \fi
    \vskip \belowcaptionskip}

\def\figstrut#1{\hbox to\linewidth{\vrule height#1\hfill}}

\makeatother

\usepackage{graphicx}

\newtheorem{notion}{Notion}{\bfseries}{\itshape}

\shtitle{The World as Evolving Information}
\title{The World as Evolving Information}
\author{Carlos Gershenson$^{1,2,3}$ 
\affil{$^{1}$ Computer Sciences Department, Instituto de Investigaciones en Matem\'aticas Aplicadas y en Sistemas\\
Universidad Nacional Aut\'onoma de M\'exico\footnote{Current affiliation. A considerable part of this work was developed while at other institutions.} \\
Ciudad Universitaria, A.P. 20-726, 01000 M\'exico D.F. M\'exico\\
cgg@unam.mx \
http://turing.iimas.unam.mx/~cgg \\
$^{2}$ New England Complex Systems Institute\\
238 Main Street Suite 319 Cambridge, MA 02142, USA\\
$^{3}$Centrum Leo Apostel, Vrije Universiteit Brussel\\
Krijgskundestraat 33 B-1160 Brussel, Belgium}}

\abstract{
This paper discusses the benefits of describing the world as information, especially in the study of the evolution of life and cognition. Traditional studies encounter problems because it is difficult to describe life and cognition in terms of matter and energy, since their laws are valid only at the physical scale. However, if matter and energy, as well as life and cognition, are described in terms of information, evolution can be described consistently as information becoming more complex.

The paper presents eight tentative laws of information, valid at multiple scales, which are generalizations of Darwinian, cybernetic, thermodynamic, psychological, philosophical, and complexity principles. These are further used to discuss the notions of life, cognition and their evolution.
}

\begin{document}

\maketitle

\section{Introduction}

Throughout history we have used concepts from our current technology as metaphors to describe our world. Examples of this are the description of the body as a factory during the Industrial Age, and the description of the brain as a computer during the Information Age. These metaphors are useful because they extend the knowledge acquired by the scientific and technological developments to other areas, illuminating them from a novel perspective. For example, it is common to extend the particle metaphor used in physics to other domains, such as crowd dynamics \cite{HelbingVicsek1999}. Even when people are not particles and have very complicated behaviour, for the purposes of crowd dynamics they can be effectively described as particles, with the benefit that there is an established mathematical framework suitable for this description. Another example can be seen with cybernetics \cite{Ashby1956,HeylighenJoslyn2001}, where the system metaphor is used: everything is seen as a system with inputs, outputs, and a control
that regulates the internal variables of the system under the influence of perturbations from its environment. Yet another example can be seen with the computational metaphor \cite{Wolfram2002}, where the universe can be modelled with simple discrete computational machines, such as cellular automata or Turing machines.

Having in mind that we are using metaphors, this paper proposes to extend the concept of information to describe the world: from elementary particles to galaxies, with everything in between, particularly life and cognition. There is no suggestion on the nature of reality as information \cite{Wheeler1990}. This work only explores the advantages of \emph{describing} the world as information. In other words, there are no ontological claims, only epistemological.

In the next section, the motivation of the paper is presented, followed by a section describing the notion of information to be used throughout the paper. In Section \ref{s:laws}, eight tentative laws of information are put forward. These are applied to the notions of life (Section \ref{s:life}) and cognition (Section \ref{s:cog}). The paper closes presenting future work and conclusions.

\section{Why Information?}

There is a great interest in the relationship between energy, matter, and information \cite{Kauffman2000,Umpleby2004,MorowitzSmith2006}. One of the main reasons for this arises because this relationship plays a central role in the definition of life: Hopfield \cite{Hopfield1994} suggests that the difference between biological and physical systems is given by the meaningful information content of the former ones. Not that information is not present in physical systems, but---as Roederer puts it---information is \emph{passive} in physics and \emph{active} in biology \cite{Roederer2005}. However, it becomes complicated to describe how this information came to be in terms of the physical laws of matter and energy. In this paper the inverse approach is proposed: let us describe matter and energy in terms of information. If atoms, molecules and cells are described as information, there is no need of a \emph{qualitative} shift (from non-living to living matter) while describing the origin and evolution of life: this is translated into a \emph{quantitative} shift (from less complex to more complex information). 

There is a similar problem when we study the origin and evolution of cognition \cite{Gershenson2004}: it is not easy to describe cognitive systems in terms of matter and energy. The drawback with the physics-based approach to the studies of life and cognition is that it requires a new category, that in the best situations can be referred to as ``emergent". Emergence is a useful concept, but it this case it is not explanatory. Moreover, it stealthily introduces a dualist view of the world: if we cannot relate properly matter and energy with life and cognition, we are forced to see these as separate categories. Once this breach is made, there is no clear way of studying or understanding how systems with life and cognition evolved from those without it. If we see matter and energy as particular, simple cases of information, the dualist trap is avoided by following a continuum in the evolution of the universe. Physical laws are suitable for describing phenomena at the physical scale. The tentative laws of information presented below aim at being suitable for describing phenomena \emph{at any scale}. Certainly, there are other approaches to describe phenomena at multiple scales, such as general systems theory and dynamical systems theory. These approaches are not exclusive, since one can use several of them, including information, to describe different aspects of the same phenomena.

Another benefit of using information as a basic descriptor for our world is that the concept is well studied and formal methods have already been developed \cite{CoverThomas2006,ProkopenkoEtAl2007}, as well as its philosophical implications have been discussed \cite{Floridi2003}. Thus, there is no need to develop a new formalism, since information theory is well established. I borrow this formalism and interpret it in a new way.

Finally, information can be used to describe other formalisms: not only particles and waves, but also systems, networks, agents, automata, and computers can be seen as information. In other words, it can contain other descriptions of the world, potentially exploiting their own formalisms. Information is an \emph{inclusive} formalism.

\section{What Is Information?}

Extending the notion of Umwelt \cite{vonUexkull1957}, the following notion of information can be given: 

\begin{notion}
\label{notion:Info}
Information is anything that an agent can sense, perceive, or observe.
\end{notion}

This notion is in accordance with Shannon's \cite{Shannon1948}, where information is seen as a just-so arrangement, a defined structure, as opposed to randomness \cite{Cohen2000,Cohen2006}, and it can be measured in bits.
This notion can be applied to everything that surrounds us, including matter and energy, since we can perceive it---because it has a defined structure---and we are agents, according to the following notion:

\begin{notion}
\label{notion:Agent}
An agent is a description of an entity that \emph{acts} on its environment \cite[p. 39]{GershensonDCSOS}.
\end{notion}

Noticing that agents (and their environments) are also information (as they can be perceived by other agents, especially us, who are the ones who \emph{describe} them as agents), an agent can be a human, a cell, a molecule, a computer program, a society, an electron, a city, a market, an institution, an atom, or a star. Each of these can be described (by us) as \emph{acting} in their environment, simply because they \emph{interact} with it.
However, not all information is an agent, e.g. temperature, color, velocity, hunger, profit.

\begin{notion}
\label{notion:Environment}
The environment of an agent consists of all the information \emph{interacting} with it.
\end{notion}

Information will be relative to the agent perceiving it\footnote{Shannon's information \cite{Shannon1948} deals only with the technical aspect of the transmission of information and not with its \emph{meaning}, i.e. it neglects the semantic aspect of communication.}. Information can exist in theory ``out there", independently of an agent, but for practical purposes, it can be only spoken about once an agent---not necessarily a human---perceives / interacts with it. The \emph{meaning} of the information will be given by the \emph{use} the agent perceiving it makes of it \cite{Wittgenstein1999}, i.e. how the agent responds to it \cite{AtlanCohen1998}. Thus, Notion \ref{notion:Info} is a \emph{pragmatic} one. Note that perceived information is different from the meaning that an agent gives to it. Meaning is an \emph{active} product of the \emph{interaction} between information and the agent perceiving it \cite{Cohen2006,Neuman:2008}.

Like this, an electron can be seen as an agent, which perceives other electrons as information. The same description can be used for molecules, cells, and animals. We can distinguish:

\begin{description}
\item[First order information] is that which is perceived directly by an agent. For example, the information received by a molecule about another molecule
\item[Second order information] is that which is perceived by an agent about information perceived by another agent. For example, the information perceived by a human observer about a molecule receiving information about another molecule.
\end{description}

Most of the scientific descriptions about the world are second order information, as we perceive how agents perceive and produce information. The present approach also introduces naturally the role of the observer in science, since everything is ``observing" the (limited, first order) information it interacts with from its own perspective. Humans would be second-level observers, observing the information observed by information. Everything we can speak about is observed, and all agents are observers.

Information is not necessarily conserved, i.e. it can be created, destroyed, or transformed. These can take place only through interaction. \emph{Computation} can be seen as the \emph{change} in information, be it creation, destruction, or transformation.
Matter and energy can be seen as particular types of information that cannot be created or destroyed, only transformed, along with the well-known properties that characterize them.

The amount of information required to describe a process, system, object, or agent determines its \emph{complexity} \cite{ProkopenkoEtAl2007}. According to our current knowledge, during the evolution of our universe there has been a shift from simple information towards more complex information \cite{Adami2002} (the information of an atom is less complex than that of a molecule, than that of a cell, than that of a multicellular organism, etc.). This ``arrow of complexity"\cite{Bedau1998} in evolution can guide us to explore general laws of information.

\section{Tentative Laws of Information}
\label{s:laws}

Seeing the world as information allows us to describe general laws that can be applied to everything we can perceive. Extending Darwin's theory \cite{Darwin1998}, the present framework can be used to reframe ``universal Darwinism" \cite{Dennet1995}, which explores the idea of evolution beyond biological systems.
In this work, the laws that describe the general behaviour of information as it evolves are introduced. These laws are only \emph{tentative}, in the sense that they are only presented with arguments in favour of them, but they still need to be thoroughly tested.

\subsection{Law of Information Transformation}
%or variation???

Since information is relative to the agents perceiving it, \emph{information will potentially be \emph{transformed} as different agents perceive it}. Another way of stating this law is the following: \emph{information will potentially be transformed by \emph{interacting} with other information}.
This law is a generalization of the Darwinian principle of random variation, and ensures \emph{novelty} of information in the world. Even when there might be static information, different agents can perceive it differently and interact with it, potentially transforming it.
Through evolution, the transformation of information generates a \emph{variety} or \emph{diversity} that can be used by agents for novel purposes.

Since information is not a conserved quantity, it can increase (created), decrease (destroyed), or be maintained as it is transformed.

As an example, RNA polymerase (RNAP) can make errors while copying DNA onto RNA strands. This slight random variation can lead to changes in the proteins for which the RNA strands serve as templates. \emph{Some} of these changes will lead to novel proteins that might improve or worsen the function of the original proteins.

The transformation of information can be classified as follows:

\begin{description}
\item[Dynamic.] Information changes itself. This could be considered as ``objective, internal" change.
\item[Static.] The agent perceiving the information changes, but the information itself does not change. There is a dynamic change but in the agent. This could be considered as ``subjective, internal" change.
\item [Active.] An agent changes information in its environment. This could be considered as an ``objective, external" change.
\item[Stigmergic.] An agent makes an active change of information, which changes the perception of that information by another agent. This could be considered as ``subjective, external" or ``intersubjective" change.
\end{description}

%***could add a table of classification

\subsection{Law of Information Propagation}

\emph{Information \emph{propagates} as fast as possible}. Certainly, only some  information manages to propagate. In other words, we can assume that different information has a different ``ability" to propagate, also depending on its environment. The ``fitter" information, i.e. that which manages to persist and propagate faster and more effectively, will prevail over other information. This law generalizes the Darwinian principle of natural selection, the maximum entropy production principle \cite{MartyushevSeleznev2006} (entropy can also be described as information), and Kauffman's tentative fourth law of thermodynamics\footnote{``The workspace of the biosphere expands, on average, as fast as it can in this coconstructing biosphere" \cite[p. 209]{Kauffman2000}}. It is interesting that this law contains the second law of thermodynamics, as atoms interact, propagating information homogeneously. It also describes living organisms, where genetic information is propagated across generations. And it also describes cultural evolution, where information is propagated among individuals. Life is ``far from thermodynamic equilibrium" because it constrains \cite{Kauffman2000} the (more simple) information propagation at the thermodynamic scale, i.e. the increase of entropy, exploiting structures to propagate (or maintain) the (more complex) information at the biological scale.

In relation with the law of information transformation, as information requires agents to perceive it, information will be potentially transformed. This source of novelty will allow for the ``blind" exploration of better ways of propagating information, according to the agents perceiving it and their environments.

Extending the previous example, if errors in transcription made by RNAP are beneficial for its propagation (which entails the propagation of the cell producing RNAP), cells with such novel proteins will have better chances of survival than their ``cousins" without transcription errors.

The propagation of information can be classified as follows:

\begin{description}
\item[Autonomous.] Information propagates by itself. Strictly speaking, this is not possible, since at least some information is determined by the environment. However, if more information is produced by itself than by its environment, we can call this autonomous propagation (See Section \ref{s:life}). 
\item[Symbiotic. ] Different information cooperates, helping to propagate each other.
\item[Parasitic. ] Information exploits other information for its own propagation. 
\item[Altruistic. ] Information promotes the propagation of other information at the cost of its own propagation.
\end{description}

\subsection{Law of Requisite Complexity}

Taking into account the law of information transformation, transformed information can increase, decrease, or maintain its previous complexity, i.e. amount \cite{ProkopenkoEtAl2007}. However, \emph{more complex information will require more complex agents to perceive, act on, and propagate it}. This law generalizes the cybernetic law of requisite variety \cite{Ashby1956}. Note that simple agents can perceive and interact with \emph{part} of complex information, but they cannot (by themselves) propagate it. An agent cannot perceive (and thus contain) information more complex than itself. For simple agents, information that is complex for us will be simple as well. As stated above, different agents can perceive the same information in different ways, giving it different meanings.

The so called ``arrow of complexity" in evolution \cite{Bedau1998} can be explained with this law. If we start with simple information, its transformation will produce by simple drift \cite{McShea1996,Miconi:2008} increases in the complexity of information, without any goal or purpose. This occurs simply because there is an open niche for information to become more complex as it varies. But this also promotes agents to become more complex to exploit novel (complex) information and propagate it. Evolution does not need to favour complexity in any way: information just propagates to every possible niche as fast as possible, and it seems that there is often an ``adjacent possible" \cite{Kauffman2000} niche of greater complexity.

%Symmetry breaking in a system marks an increase of its complexity, since a symmetric configuration is less complex than an asymmetric one.

For example, it can be said that a protein (as an agent) perceives some information via its binding sites, as it recognizes molecules that ``fit" a site. More complex molecules will certainly need more complex binding sites. Whether complex molecules are better or worse is a different matter: some will be better, some will be worse. But for those which are better, the complexity of the proteins must match the complexity of the molecules perceived. If the binding site perceives only a part of the molecule, then this might be confused with other molecules which share the perceived part. Following the law of information transformation, there will be a variety of complexities of information. The law of requisite complexity just states that the increase in complexity of information is determined by the ability of agents to perceive, act on, and propagate more complex information.

Since more complex information will be able to produce more variety, the \emph{speed} of the complexity increase will escalate together with the complexity of the information.

\subsection{Law of Information Criticality}

\emph{Transforming and propagating information will tend to a critical \emph{balance} between its stability and its variability}. Propagating information  maintains itself as much as possible, but transforming information varies it as much as possible. This struggle leads to a critical balance analogous to the ``edge of chaos" \cite{Langton1990,Kauffman1993}, self-organized criticality \cite{BTW1987,Adami1995}, and the ``complexity from noise" principle \cite{Atlan1974}. The homeostasis of living systems can also be seen as the self-regulation of information criticality. 

This law can generalize Kauffman's four candidate laws for the coconstruction of a biosphere \cite[Ch. 8]{Kauffman2000}. Their relationship with this framework demands further discussion, which is out of the scope of this paper.

A well known example can be seen with cellular automata \cite{Langton1990} and random Boolean networks \cite{Kauffman1993,Gershenson2004c,Gershenson:2010}: stable (ordered) dynamics limit considerably or do not allow change of states so information cannot propagate, while variable (chaotic) dynamics change the states too much, losing information.
Following the law of information propagation, information will tend to a critical state between stability and variability to maximize its propagation: if it is too stable, it will not propagate, and if it is too variable, it will be transformed. In other words, ``critical" information will be able to propagate better than stable or variable one, i.e. as fast as possible (cf. law of information propagation).

\subsection{Law of Information Organization}

\emph{Information produces constraints that regulate information production}. These constraints can be seen as \emph{organization} \cite{Kauffman2000}. In other words, evolving information will be organized (by transformation and propagation) to regulate information production. According to the law of information criticality, this organization will lie at a critical area between stability and variability. And following the law of information propagation, the organization of information will enable it to propagate as fast as possible.

This law can also be seen as information having a certain \emph{control} over its environment, since the organization of information will help it withstand perturbations. It has been shown \cite{KlyubinEtAl2004,ProkopenkoEtAl2006,KlyubinEtAl2007} that using this idea as a fitness function can lead to the evolution of robust and adaptive agents, namely maximizing the mutual information between sensors and environment.

A clear example of information producing its own organization can be seen with living systems, which are discussed in Section \ref{s:life}.

%***thermodynamics???

%***MST? satisfaction/friction???

\subsection{Law of Information Self-organization}

\emph{Information tends to its preferred, most probable state}. This is actually a tautology, since observers determine probabilities after observing tendencies of information dynamics. Still, this tautology can be useful to describe and understand phenomena. This law lies at the heart of probability theory and dynamical systems theory \cite{Ashby1962}. The dynamics of a system tend to a subset of its state space, i.e. attractors, depending on its history. This simple fact reduces the possibility space of information, i.e. a system will tend towards a small subset of all possible states. If we describe attractors as ``organized", then we can describe the dynamics of information in terms of self-organization \cite{GershensonHeylighen2003a}.

Pattern formation can be described as information self-organizing, and related to the law of information propagation. Information will self-organize in ``fit" patterns that are the most probable (defined \emph{a posteriori}).

Understanding different ways in which self-organization is achieved by transforming information can help us understand better natural phenomena \cite{Gershenson:2010a} and design artificial systems \cite{GershensonDCSOS}.  For example, random Boolean networks can be said to self-organize towards their attractors \cite{Gershenson:2010}.

\subsection{Law of Information Potentiality}

\emph{An agent can give different potential meanings to information}. This implies that the same information can have different meanings. Moreover, meaning---while being information---can be independent of the information carrying it, i.e. depend only on the agent observing it. Thus, different information can have the same potential meaning. The precise meaning of information will be given by an agent observing it within a specific context.

The potentiality of information allows the effective communication between agents. Different information has to be able to acquire the same meaning (homonymy), while the same information has to be able to acquire different meanings (polysemy) \cite{Neuman:2008}. The relationship between the laws of information and communication is clear, but beyond the scope of this paper.

The law of information potentiality is related to a passive information transformation, i.e. a change in the agent observing information.

In spite of information potentiality, not all meanings will be suitable for all information. In other words, pure subjectivism cannot dictate meanings of information. By the law of information propagation, some meanings will be more suitable than others and will propagate. The suitability of meanings will be determined by their use and context \cite{Wittgenstein1999}. However, there is always a certain freedom to subjectively transform information. 

For example, a photon can be observed as a particle, as a wave, or as a particle-wave. The suitability of each given meaning is determined by the context in which the photon is described/observed.

\subsection{Law of Information Perception}

\emph{The meaning of information is \emph{unique} for an agent perceiving it in unique, always changing open contexts}. If meaning of information is determined by the use an agent makes of it, which is embedded in an open environment, we can go to such a level of detail that the meaning will be unique. Certainly, agents make generalizations and abstractions of perceptions in order to be able to respond to novel information. Still, the precise situation and context will never be repeated. This makes perceived information unique. The implication of this is that the response to any given information might be ``unexpected", i.e. novelty can arise. Moreover, the meaning of information can be to a certain extent \emph{arbitrary}. This is related with the law of information transformation, as the uniqueness of meaning allows the same information perceived differently by the same or different agents to be statically transformed. 

This law is a generalization of the first law of human perception: ``whatever is perceived can be perceived only from a uniquely situated place in the overall structure of points of view" \cite[p. xxiv]{Holquist:1990} (cited in \cite[p. 250]{Neuman:2008}). We can describe agents perceiving information as filtering it. An advantage of humans and other agents is that we can \emph{choose} which filter to use to perceive. The suggestion is not that ``unpleasant" information should be solipsistically ignored, but that information can be potentially actively transformed.

For example, T lymphocytes in an immune system can perceive foreign agents and attack them. Even when the response will be similar for similar foreign agents, each perception will be unique, a situation that always leaves space for novelty.

\subsubsection{Scales of perception}

Different information is perceived at different scales of observation \cite{BarYam2004}. As the scale tends to zero, then the information tends to infinite. For lower scales, more information and details are perceived. The uniqueness of information perception dominates at these very low (spatial and temporal) scales. However, as generalizations are made, information is ``compressed", i.e. only relevant aspects of information are perceived\footnote{The relevance is determined by the context, i.e. different aspects will be relevant for different contexts.}. At higher scales, more abstractions and generalizations are made, i.e. less information is perceived. When the scale tends to infinite, the information tends to zero. In other words, no information is needed to describe all of the universe, because all the information is already there. This most abstract understanding of the world is in line with the ``highest view" of Vajrayana Buddhism \cite{Nydahl:2008}. Implications at this level of description cannot be right or wrong, because there is no context. Everything is contained, but no information is needed to describe it, since it is already there. This ``maximum" understanding is also described as vacuity, which leads to bliss \cite[p. 42]{Nydahl:2008}.

%***graph? equation?

Following the law of information criticality, agents will tend to a balance where the perceived information is minimal but maximally predictive \cite{Shalizi2001} (at a particular scale): few information is cheaper, but more information in general entails a more precise predictability. The law of requisite complexity applies at particular scales, since a change of scale will imply a change of complexity of information \cite{BarYam2004}.

\section{On the Notion of Life}
\label{s:life}

There is no agreed notion of life, which reflects the difficulty of defining the concept. Still, many researchers have put forward properties that characterize important aspects of life. \emph{Autopoiesis} is perhaps the most salient one, which notes that living systems are self-producing \cite{VarelaEtAl1974,McMullin2004}. Still, it has been argued that autopoiesis is a necessary but not sufficient property for life \cite{RuizMoreno2004}. The relevance of autonomy \cite{Barandarian2004,MorenoRuiz2006,KrakauerZanotto2007} and individuality \cite{Michod2000,KrakauerZanotto2007} for life have also been highlighted . 

These approaches are not unproblematic, since no living system is completely autonomous. This follows from the fact that all living systems are open. For example, we have some degree of autonomy, but we are still dependent on food, water, oxygen, sunlight, bacteria living in our gut, etc. This does not mean that we should abandon the notion of autonomy in life. However, we need to abandon the sharp distinction between life and non-life \cite{Bedau1998,KrakauerZanotto2007}, as different degrees of autonomy escalate \emph{gradually}, from the systems we considered as non-living to the ones we consider as living. In other words, life has to be a fuzzy concept.

Under the present framework, living and non-living systems are information. Rather than a yes/no definition, we can speak about a ``\emph{life ratio}":

\begin{notion}
\label{notion:Life}
The ratio of \emph{living information} is the information produced by itself over the information produced by its environment.
\end{notion}
%***equation?

Being more specific---since all systems also receive information---a system with a high life ratio produces more (first order) information about itself than the one it receives from its environment. Following the law of information organization, this also implies that living information produces more of its own constraints (organization) to regulate itself than the ones produced by its environment, and thus it has a greater autonomy. All information will have constraints from other (environmental) information, but we can measure (as second-order information) the proportion of internal over external constraints to obtain the life ratio. If this is greater than one, then the information regulates by itself more than the proportion that is regulated by external information. In the opposite case, the life ratio would be less than one.

Following the law of information propagation, evolution will tend to information with higher life ratios, simply because this can propagate better, as it has more ``control" and autonomy over its environment. When information depends more on its environment for its propagation, it has a higher probability of being transformed as it interacts with its environment.

Note that the life ratio depends on spatial and temporal scales at which information is perceived. For example, for some microorganisms observed at a scale of years , the life ratio would be less than one, but if observed at a scale of seconds, the life ration would be greater than one.

Certainly, some artificial systems would be considered as living under this notion. However, we can make a distinction between living systems embodied in or composed by biological cells \cite{DeDuve2003}, i.e. life as we know it, and the rest, i.e. life as it could be. The latter ones are precisely those explored by artificial life.
 
\section{On the Notion of Cognition}
\label{s:cog}

Cognition is certainly related with life \cite{Stewart1995}. The term has taken different meanings in different contexts, but all of them can be generalized into a common notion \cite{Gershenson2004}. Cognition comes from the Latin \emph{cognoscere}, which means ``get to know". Like this,

\begin{notion}
\label{notion:Cognition}
A system is cognitive if it \emph{knows} something \cite[p.135]{Gershenson2004}.
\end{notion}

From Notion \ref{notion:Agent}, all agents are cognitive, since they ``\emph{know}" how to act on their environment, giving (first order) \emph{meaning} to their environmental information. Thus, there is no boundary between non-cognitive and cognitive systems. Throughout evolution, however, there has been a \emph{gradual} increase in the complexity of cognition \cite{Gershenson2004}. This is because all agents can be described as possessing some form of cognition, i.e. ``knowledge" about the (first-order) information they perceive\footnote{One could argue that, since agency (and thus cognition) is already assumed in all agents, this approach is not explanatory. But I am not trying to explain the ``origins" of agency, since I assume it to be there from the start. I believe that we can only study the evolution and complexification of agency and cognition, not their ``origins".}.

Following the law of requisite complexity, evolution leads to more complex agents, to be able to cope with the complexity of their environment. This is precisely what triggers the (second-order) increase in the complexity of cognition we observe.

Certainly, there are different types of cognition\footnote{For example, human, animal, plant, bacterial, immune, biological, adaptive, systemic, and artificial \cite{Gershenson2004}.}. We can say that a rock ``knows" about gravity because it perceives its information, which has an effect on it, but it cannot \emph{react} to this information. Throughout evolution, information capable of maintaining its integrity has prevailed over that which was not. \emph{Robust} information is that which can resist perturbations to maintain its integrity. The ability to react to face perturbations to maintain information makes information \emph{adaptive}, increasing its probability of maintenance. When this reaction is made before it occurs, the information is \emph{anticipative}\footnote{For a more detailed treatment on robustness, adaptation, and anticipation, see \cite{GershensonDCSOS}}. As information becomes more complex (even if only by information transformation), the mechanisms for maintaining this information also become more complex, as stated by the law of requisite complexity. This has led gradually to the advanced cognition that animals and machines posses.

\section{Future Work}

The ideas presented here still need to be explored and elaborated further. One way of doing this would be with a simulation-based method. Being inspired by $\epsilon$-machines \cite{Shalizi2001,GoernerupCrutchfield2008}, one could start with ``simple" agents that are able to perceive and produce information, but cannot control their own production. These would be let to evolve, measuring if complexity increases as they evolve. The hypothesis is that complexity would increase (under which conditions still remains to be seen), to a point where ``$\epsilon$-agents" will be able to produce themselves depending more on their own information than that of the environment. This would be similar to the evolution in Tierra \cite{Ray1991} or Avida \cite{AdamiBrown1994} systems, only that self-replication would not be inbuilt. The tentative laws of information presented in Section \ref{s:laws} would be better defined if such a system was studied.

One  important aspect that remains to be studied is the representation of thermodynamics in terms of information. This is because the ability to perform thermodynamic work is a characteristic property of biological systems \cite{Kauffman2000}. This work can be used to generate the organization necessary to sustain life (cf. law of information organization). It is difficult to describe life in terms of thermodynamics, since it entails new characteristic properties not present in thermodynamic systems. But if we see the latter ones as information, it will be easier to describe how life---also described as information---evolves from them, as information propagates itself at different scales.

A potential application of this framework would be in economy, considering capital, goods, 
and resources as information (a non-conserved quantity) \cite{FarmerZamani2006}. A similar benefit (of non-conservation) could be given in game theory: if the payoff of games is given in terms of information (not necessarily conserved), non-zero sum games could be easier to grasp than if the payoff is given in material (conserved) goods.

It becomes clear that information (object), the agent perceiving it (subject) and the meaning-making or transformation of information (action) are deeply interrelated. They are part of the same totality, since one cannot exist without the others. This is also in line with Buddhist philosophy. The implications of an informational description of the world for philosophy have also to be addressed, since some schools have focussed on partial aspects of the object-subject-action trichotomy.  Another potential application of the laws of information would be in ethics, where value can be described accordingly to the present framework.

\section{Conclusions}

This paper introduced general ideas that require further development, extension and grounding in particular disciplines. Still, a first step is always necessary, and hopefully feedback from the community will guide the following steps of this line of research. %Here mainly philosophical issues were addressed. The explanatory and predictive benefits of this framework still remain to be addressed.

Different metaphors for describing the world can be seen as different languages: they can refer to the same objects without changing them. And each can be more suitable for a particular context. For example, English has several advantages for fast learning, German for philosophy, Spanish for narrative, and Russian for poetry. In other words, there is no ``best" language outside a particular context. In a similar way, I am not suggesting that describing the world as information is more suitable than physics to describe physical phenomena, or better than chemistry to describe chemical phenomena. It would be redundant to describe particles as information if we are studying only particles. The suggested approach is meant only for the cases when the physical approach is not sufficient, i.e. across scales, constituting an alternative worth exploring to describe evolution.

It seems easier to describe matter and energy in terms of information than vice versa. Moreover, information could be used as a common language across scientific disciplines \cite{vonBaeyer2004}.

\subsection*{Acknowledgements}

I should like to thank Irun Cohen, Inman Harvey, Francis Heylighen, David Krakauer, Antonio del R\'{i}o, Marko Rodriguez, David Rosenblueth, Stanley Salthe, Mikhail Prokopenko, Cl\'{e}ment Vidal, and H\'{e}ctor Zenil for their useful comments and suggestions. 

\footnotesize{
\bibliography{carlos,information,sos,complex,COG,rbn,evolution}
}
\end{document}